\documentclass[11pt,twoside]{article}
\font\openface=msbm10 at10pt 
\def\Minkowski {{\hbox{\openface M}}} 
\begin{document}\hbadness=10000\thispagestyle{empty}

\begin{center} 
{\LARGE {\bf Spacetime and Matter -- \\ \vskip 0.2cm a duality of partial orders}}\,\,\footnote{
Awarded fourth prize in the 2009 FQXi essay contest \emph{``What is Ultimately Possible in Physics?''}.}
\vskip 0.4cm 
{\large Hans-Thomas Elze \vskip 0.2cm Dipartimento di Fisica ``Enrico Fermi'', Universit\`a di Pisa}
\end{center}

\noindent
{\small {\bf Abstract.} A new kind of duality between the deep structures of spacetime and 
matter is proposed here, considering two partial orders which incorporate 
causality, extensity, and discreteness. This may have surprising 
consequences for the emergence of quantum mechanics, which are discussed.}
%%%%%%%%%%%%%%%%%%%%%%
\section{The context}

$\mathcal{PROLOGUE:}$ \emph{``What is Ultimately Possible in Physics?''} -- This 
question must occur to every physicist, now and then, in those rare moments when 
the work can be put aside, when there are no grant applications to review, nor are there other 
earthly problems to worry about immediately. Instead, perhaps, there is an unimpeded 
view of the Milky Way or of the ocean surf \,\ldots \\ \noindent 
In this essay, we address the question posed by successively narrowing it down. \\ \noindent 
\emph{``Will there be a} T\emph{heory} O\emph{f} E\emph{verything?''} -- This one cannot 
be answered without having the TOE. Having the TOE, \emph{we} would not 
recognize the TOE. \\ \noindent 
More modestly, \emph{``Will there be quantum gravity?''}. -- In trying to tackle this one, 
we will analyze heuristically some apparent deficiency in current elaborations of the 
theme, resulting in promising new perspectives on the penultimately possible in physics. \\ 
\begin{center} 
\begin{tabular}{p{5cm}}
\hline \phantom .
\end{tabular}
\end{center} 
\emph{``There is no quantum gravity''} -- might this author be tempted to say. 
While a majority of physicists presumably would hold on to \emph{``Not yet, but ... ''}.  

Adepts of various contenders for a theory of quantum gravity, such as string theory, 
loop quantum gravity, or quantum geometrodynamics would fill in achievements and  
problematic issues traded between these schools. 

To wit, no consensus has been reached despite  
intense study of this stumbling block on the road towards a  
unified picture of the Universe. Brilliant thinkers have tried their best, last not least, 
motivated by astounding successes of the Standard Model of the constituents of matter and 
of the forces through which they interact. 
\vskip 0.3cm
Purpose of this essay is \emph{not} to propose another surprise model 
which could subsume gravity and  
spacetime to the successful paradigm of quantum field theory. 
Such theories commonly depart from \emph{classical} Newtonian concepts applied to fields 
which represent matter (quarks, leptons, and gauge bosons), somehow existing in 
spacetime. These theories are then \emph{``quantized''}. 
Following a precise protocol, they are reformulated and generalized 
according to abstract \emph{axioms}, or derived rules, which have been     
distilled from research in quantum mechanics, as described elegantly and succinctly   
in Dirac's famous book \cite{Dirac}. 
 
With rapid advances of experimental techniques, foundational 
questions of quantum mechanics have entered center-stage in recent years. It is also 
\emph{not} the place here to discuss different related interpretations nor conceptual 
problems within the mathematical framework \cite{SchlosshauerZurekKiefer}. 
\vskip 0.3cm
Instead, we must recall a deep structural disparity between Quantum Theory (\emph{QT}) and 
the classical theory of gravity and spacetime, Einstein's General Relativity (\emph{GR}). 
  
The former describes quantum evolution 
between an initial and a final state, 
pertaining to an initial and final moment of time, respectively. The law of  
motion, e.g., in form of the Schr\"odinger equation, requires a slicing 
of spacetime according to an external time. Dynamical changes happen from slice to slice, 
while a clock ``ticks'', similarly as in 
Newtonian mechanics. This time is still universal but not absolute, since \emph{QT} 
can be adjusted to Special Relativity.  

But where is such clock to be found? -- 
Reference to anyone's time keeping device or to periodic natural phenomena may 
seem to suffice. But what about the whole Universe? It is a single entity and, therefore, 
no \emph{external} clock exists! 
 
Furthermore, modern theories of gravity and spacetime are founded on the 
principle that meaningfull physical statements must be independent of 
a choice spatial or temporal coordinates.  
This symmetry principle has changed notions of \emph{Where} and \emph{When} profoundly.  
It introduces an aspect of arbitrariness into the concept of time which,   
nevertheless, remains essential for \emph{QT}, as it is. 
 
This disparity between \emph{GR} and \emph{QT},  
produces major obstacles on the way to 
a unified spacetime-matter theory \cite{KieferRovelli}. 
It motivates our heuristic argument, which 
we outline here, with details to follow in this essay.
\vskip 0.3cm 
At the level of quantum field theory, \emph{QT} complies with 
Special Relativity. One proceeds without paying attention to the 
\emph{deep structure of spacetime}. With the exception of worries caused by ubiquitous 
infinities in such theories. 

While infinities can be dealt with 
by ``renormalization'', they do arise from phenomena at very short 
distances -- tacitly assuming that spacetime is a  
continuum manifold, admitting higher and higher energy probes that 
resolve shorter and shorter distances, in principle.  
This does not imply open ended search for ``fundamental'' 
properties of matter, its atomistic aspects, in particular. The 
very assumption that a theory of spacetime compatible with quantum mechanics does  
exist also suggests an ultimate cut-off: by \emph{QT}, 
spacetime is expected to become 
discrete at Planck scale, $l_P\approx 10^{-35}$m. 

Spacetime, therefore, should reflect a similar (or the same?) \emph{atomism} 
as attributed to matter in modern physical theories. 
Such ideas can be traced back all the way to ancient Greece and  
the pre-Socratic philosophers, as described, for example, in Schr\"odinger's marvellous  
book \cite{Schroedinger}. Problematic aspects of space as continuum have already been 
noticed then, with much insight, as well as in later periods when 
natural philosophy thrived \cite{Jammer,SorkinLecture}.   
\vskip 0.3cm
This leads us to conclude our brief exposition of ideas addressing  
``spacetime and matter'' with the \emph{hypothesis}: 
\begin{itemize}
\item Spacetime and matter must reflect each others atomistic structure.    
\end{itemize}
Hence, unequal footing for matter and spacetime may cause obstructions to the  
search for ``quantum gravity'' from the outset. 
Of course, this results from our ignorance 
of the relevant degrees of freedom close to $l_P$, where quantum numbers and degrees of 
freedom of the Standard Model not necessarily play a role.

Furthermore, spacetime without matter or matter in a ``background'' spacetime  
are justifiable as approximate yet accurate mathematical models 
under certain circumstances. However, these  
abstractions found their way into the discussed disparity between \emph{GR} and \emph{QT}.

Therefore, we will address the fundamental mutual dependence of spacetime and matter, 
and its consequences. We assume that the categories \emph{causality}, 
embedded in the deepest spacetime structure, and \emph{extensity},  
a defining quality of matter here, are \emph{dual} to each other, in a sense 
to be defined. Physical reality ultimately rests inseparably on 
both.~\footnote{This author has a hard time to imagine a piece of matter to exist 
without an accompanying spacetime volume and finds it just as difficult to believe 
that a spacetime can exist deprived of matter, which provides rulers and light signals  
to reveil its geometry.}   

%%%%%%%%%%%%%%%%%%%%%%%%%%%%%%%%%%%%%%%%%%%%%%%%%%%%%%%%%%%%%%%%%%%%%%%%%
\section{Atomism all the way -- \\ locally finite partially ordered sets} 
A \emph{locally finite partially ordered set} is mathematically defined as 
a set $C$ together with a binary relation $\prec$ satisfying: 
\begin{itemize} 
\item \emph{transitivity}: $x\prec y\prec z\;\Rightarrow\;x\prec z\;,\;\;\forall x,y,z\in C\;$;
\item \emph{irreflexivity}: $x\not\prec x\;,\;\;\forall x\in C\;$; 
\item \emph{local finiteness}: $\forall x,z\in C$ the set $\{ y\in C|x\prec y\prec z\}$ has 
a finite number of elements. 
\end{itemize} 
A relation between two elements that is \emph{not} implied by transitivity is 
called a \emph{link}. 

Such structures made their entrance into physics only recently, while 
having much older roots in related atomistic ideas \cite{Schroedinger,Jammer,SorkinLecture}. 
Myrheim \cite{Myrheim} 
and 't\,Hooft \cite{tHooft} 
initiated modern considerations of discrete sets as  
foundation for a theory of spacetime, incorporating also causality as a fundamental principle.  
This subsequently led to the \emph{causal set hypothesis} \cite{Sorkin1} and research program; 
among the most interesting results has been a \emph{prediction} of the cosmological constant  
\cite{SorkinLecture}.~\footnote{For the state of the art, see  
papers by Rideout and Wallden, 
Sorkin, Sverdlov and Bombelli, Henson, Johnston, 
Philpott with Dowker and Sorkin, Brightwell with Henson, and Surya  
collected in \cite{DICE}.}  

No attempts to merge such discrete structures with \emph{QT}  
have been reported. In fact, it may be wrong to pursue this, since   
quantum phenomena possibly \emph{emerge} only at larger scales than $l_P$, 
the scale of spacetime discreteness.   
We expand on this issue in Section 3.

%%%%%%%%%%%%%%%%%%%%%%%%%%%%%%%%%%%%%%
\subsection{Spacetime as a causal set}
The binary relation $\prec$ of a locally 
finite partially ordered set can be interpreted as stating  
\emph{causal order} between two elements $x,y\in C$: 
\begin{equation}\label{precedes}
x\prec y\;\;\mbox{reads as}\;\; x\;\;\mbox{precedes}\;\; y
\;\;.\end{equation} 
Elements of the set are naturally interpreted as elementary ``events'' 
and $C$ is termed a \emph{causal set} or \emph{causet}. -- 
Elements which are not related by $\prec$ are \emph{spacelike} 
to each other, which is denoted by: 
$x\natural y\;\Leftrightarrow\; x\not\prec y\;\mbox{and}\; y\not\prec x\;$. 

Transitivity and irreflexivity together imply that 
there can be \emph{no causal loops}. Furthermore,  
local finiteness introduces the notion of discreteness, 
when this definition is applied to (re)construct spacetime. 

The motivation for a causet as the structure possibly underlying 
spacetime stems from the fact that causal ordering together with a four-dimensional 
volume element is sufficient to determine all metric information, topology, and 
differentiable structure of a continuum spacetime, under plausible technical 
assumptions. This has been stated as \cite{SorkinLecture}: 
\begin{equation}\label{Geom}
\mbox{Order}\;\; +\;\;\mbox{Number}\;\; =\;\;\mbox{Geometry}
\;\;,\end{equation} 
where numbers of set elements encode volume information in the discrete case. --   
Thus, Minkowski space, $\Minkowski$, can be reconstructed 
as continuum limit of a causet. 

Generally, a representative causet can always be obtained by a 
Lorentz invariant Poisson process (\emph{``sprinkling''}): elements of the set are 
generated randomly by sprinkling points with uniform average density of $1/l_P^{\; 4}$ 
into a Lorentzian four-dimensional continuum manifold, from which they inherit causal 
ordering.~\footnote{It has been conjectured that this is the only way to 
produce a Lorentz invariant discretization.}  

Dynamics of causets will not be further discussed. It is, of course, an  
important topic to take the 
stringent requirements of general coordinate invariance and causality into account, 
since evolving  
discrete structures are to replace dynamical continuum manifolds governed by {\it GR} 
at large \cite{SorkinLecture,DICE}. It may suffice to mention that through the 
growth of a causet by new elements time can be seen to ``happen''.  

However, another question looms, as surmised in Section 1, and shall be addressed: 
Has matter been overlooked in recent attempts to formulate a theory of spacetime valid 
down to shortest distances? 

%%%%%%%%%%%%%%%%%%%%%%%%%%%%%%%%%
\subsection{Matter and extensity}  
A causet correlate of spacetime is well defined \emph{without}  
mentioning matter. For physics, however, matter has to be dealt 
with. It is 
necessary, in order to derive phenomenological consequences of the atomistic 
structure of spacetime, which one would hope to test experimentally.  

So far, this has been done by adapting continuum notions of propagation of 
particles or fields to terms of ``order and number'' which are available 
for a causet; cf. Philpott, Dowker and Sorkin or Sverdlov and Bombelli 
in \cite{DICE}. 

Still, this comes with 
assumptions which detract from simplicity and beauty of the 
causal set hypothesis: 
\begin{itemize}  
\item Causet elements have to  
\emph{register} additional physical data, such as a field amplitude 
or presence of a particle, and causal relations between them must \emph{act  
as transport channels} for these quantities or related information.  
\item This might signify that such causet  
is a \emph{coarse-grained} version of a finer structure, 
with manifestations of matter \emph{emerging} together with the 
causet. 
\end{itemize} 
The latter would be in the spirit of Kaluza-Klein models, with additional 
degrees of freedom arising from the geometry of extra-dimensions beyond 
the perceived four-dimensional spacetime. It cannot be ruled out 
at present. 
\vskip 0.3cm
However, as a radical alternative, we introduce a \emph{locally finite partial order} 
to describe the presence of matter as follows. 

Lacking a general dynamical theory, consider for a moment 
that matter is positioned or moving relative to a \emph{given} causet.  
It may be naturally allocated ``in between'' causet elements 
that are \emph{spacelike} to each other. To this end, we choose a locally 
finite partial order that incorporates the category of 
\emph{``extensity''}, in distinction to \emph{causality}. 

The corresponding binary relation between spacelike elements,  
$u,v\in C$, with $u\natural v$, is denoted by $\prec^e$ and interpreted by: 
\begin{equation}\label{precedese}
u\prec^e v\;\;\mbox{reads as}\;\; u\;\;\mbox{is extended by}\;\; v
\;\;.\end{equation} 
Extensity here is a fundamental feature of matter which, in its primordial form, 
is represented by the extensional relation between events. 

%%%%%%%%%%%%%%%%%%%%%%%%%%%%%%%%%%%%%
\subsection{Spacetime-matter duality}
Oriented loops of extensional links also do \emph{not} exist. This does not rule 
out all-spacelike extended matter, including tilings of Euclidean space and loops 
in general.  
  
By definition, a causal link between two events \emph{cannot} coincide with an extensional link. 
In this sense, both orders are complementary to each other. 

This leads us to consider 
\emph{transformations} of the causal and 
extensional relations among events according to one of the    
rules: 
\begin{eqnarray}\label{A}  
A&:=&\Big ((\prec ,\succ ,\prec^e,\succ^e)\;\longmapsto\;(\prec^e,\succ^e,\prec ,\succ )\Big ) 
\;\;, \\ [1ex] \label{B} 
B&:=&\Big ((\prec ,\succ ,\prec^e,\succ^e)\;\longmapsto\;(\succ^e,\prec^e,\succ ,\prec )\Big ) 
\;\;, \\ [1ex] \label{C} 
C&:=&\Big ((\prec ,\succ ,\prec^e,\succ^e)\;\longmapsto\;(\succ ,\prec ,\succ^e,\prec^e )\Big ) 
\;\;. \end{eqnarray} 
Representing each map by a $4\times 4$ matrix, it is easy 
to see that they form an abelian group, with multiplication rules: 
$AB=C$, $BC=A$, $CA=B$, $A^2=B^2=C^2=e$, $e$ denoting 
the identity. It is known as \emph{Klein four-group} {\bf V}.~\footnote{The 
symmetric traceless matrices representing $A$, $B$, and $C$ can  
serve as generators of a continuous group. Transformations effected 
by its elements can produce \emph{superpositions} of spacetime-matter configurations. 
We do not investigate this interesting possibility here.}  
  
Consider a causet sprinkled into four-dimensional Minkowski 
space, $\Minkowski^4$, with extensional relations representing matter. Then,
global application of rule $C$ can be interpreted as discrete analogue  
of combined \emph{time reversal} and \emph{parity} transformations, 
under which such a set may or may not be invariant.

A causet can only be \emph{globally invariant} under 
(either) transformation 
$A$ (or $B=CA=AC$), if the directed network of causal and 
extensional links can be mapped one-to-one, necessitating equal   
numbers of causal and extensional links. 
Elements which are neither causally nor extensionally related to 
any other element are not affected. 
Such a highly symmetric state is characterized by \emph{spacetime-matter duality}:  
causal and extensional structures imply each other.     
   
At first sight, this duality seems artificial.   
Looking outside, local features of $\Minkowski^4$ with matter could not 
deviate more from duality between two partial orders.   
However, this is due to asymmetric circumstances with spacetime appearing   
static while matter evolves with respect to it.   

The motion of this \emph{dilute matter} can be 
described simply. Similarly to  
propagation of a scalar field from a point source \cite{DICE},   
the ``endpoint'' vertices of an extensional link propagate forward, 
each into its ``future lightcone'': all outgoing causal links 
are followed; however, only spacelike pairs of new endpoints 
contribute, which allow and consequently obtain an   
extensional link between them.~\footnote{Such propagation on a causet sprinkled 
into $\Minkowski^4$ can be analyzed in detail. If limited to overlapping future 
lightcones of the endpoints, this leads to a \emph{formation time} effect.} 

For \emph{dense matter}, we consider  
a $\Minkowski^2$ background. A corresponding causet set inherits causal and spacelike 
relations through sprinkling from $\Minkowski^2$. The latter define extensional 
links, with directionality given by the one-dimensional spatial order of the 
backgound. This discrete spacetime-matter structure is {\it invariant} 
under transformation $A$, hence \emph{selfdual} (apart from statistical fluctuations). 
It is an extended matter 
distribution resembling the world sheet of a straight line string. 
By thinning the distribution of extensional links, the amount of matter 
could be locally varied.  
\vskip 0.3cm 
How can the duality be realized in \emph{three spatial dimensions}?  
 
All sets are locally finite with denumerable elements and  
the expected number of events per unit fourvolume of spacetime is finite. 
Therefore, discretized patches of  
higherdimensional spacetime-matter can be built by randomly stacking 
(1+1)-dimensional sheets together, eventually reaching the continuum limit. 
This implies \emph{dimensional reduction} of spacetime from 3+1 to 1+1 dimensions, 
as shorter and shorter distances 
are probed. Indeed, there are various independent arguments supporting    
this picture \cite{Carlip09}. -- 
A macroscopic patch of $\Minkowski^4$ (with matter) \emph{can} result, 
if spatial orientations of the microscopic (extensional) sets are   
randomized, while keeping lightcones aligned.
Clearly, it is not trivial to arrive 
at the Poincar\'e symmetry. Predictable deviations would be most 
interesting.~\footnote{Extensional ordering introduces 
a spacelike correlation, if  
stringlike structures thread spacetime, as is implicit here. 
It might have interesting consequences 
for quantum mechanics.} 
\vskip 0.3cm 
Finally, consider any element of a causet discretizing 
$\Minkowski^4$ and take its \emph{``lightcone''},    
consisting of causally related elements. The lightcone becomes  
an extensionally ordered \emph{``matter cone''} by transformation $A$. 
The extensionally unrelated \emph{outside} of the matter cone 
can be identified with the causally unrelated outside of the lightcone, 
since $A$ does not act on unrelated elements.  
Statistical properties of these sets are not influenced by 
the transformation. Therefore, dimension measures will indicate a  
four-dimensional space in both cases \cite{SorkinLecture}. 
  
Here, the reader must wonder, whether we are only changing names between   
causal and extensional order. -- Without \emph{interactions}, 
this must be so. However, as soon as interactions affect causal and matter links, 
or vertices, in distinctive ways, these entities obtain distinctive roles, differentiating 
aspects of matter from those of spacetime.     
 
A theory with interactions phrased in terms of the partial orders awaits elaboration. 
However, duality of spacetime and matter here already  
suggests that a Lorentzian space \emph{together} with its dual should be relevant    
for physics after sufficient coarse-graining of sufficiently large 
partially ordered sets. 
Known phenomena may then take place as perturbations of perfect spacetime-matter 
duality, rendering physical a \emph{doubled number of dimensions}.  

%%%%%%%%%%%%%%%%%%%%%%%%%%%%%%%%%%%%%%%%%%%%%
\section{Emergent quantum mechanical aspects}
We have argued for duality between 
spacetime and matter in the foregoing, suggesting also  
that physics at large scales must refer to \emph{two} copies, say, 
of $\Minkowski^4$.  

Indeed, we have encountered this doubling before. -- Classical Hamiltonian mechanics employs 
phase space, consisting of coordinates \emph{paired} with momenta. 
The latter can be replaced by a second set of coordinates through \emph{Fourier transformation}. 
Both sets describe $\Minkowski^4$ spaces in Special Relativity.   

The author has recently studied classical ensemble theory in terms of such 
\emph{doubled set of coordinates}; see Ref.\,\cite{I09} with 
numerous related references. We summarize here how this, quite surprisingly, produces  
important aspects of quantum theory (\emph{QT}). 
\vskip 0.3cm
Consider a (1+1)-dimensional nonrelativistic object with equations of motion 
derived from a Hamiltonian function, $H(x,p):=p^2/2+v(x)$, where $x$, $p$, and $v$ 
denote coordinate, momentum, and \emph{true potential}, 
respectively.~\footnote{The following works   
for Lagrangian field theories as well.} -- The 
Liouville equation describes evolution of a \emph{statistical ensemble} 
of such objects by its evolving phase space probability distribution, $f$, 
with dynamics given by $H$. 
Combining Fourier transformation and linear 
coordinate transformations, the \emph{classical} equation \emph{is}:  
\begin{eqnarray}\label{Schroed} 
i\partial_tf(x,y;t)&=&\Big\{ \hat H_x-\hat H_y+{\cal E}(x,y)\Big\}f(x,y;t)
\;\;, \\ [1ex] \label{HX} 
\hat H_\chi &:=&-\frac{1}{2}\partial_\chi ^{\;2}+v(\chi )\;\;, 
\;\;\;\mbox{for}\;\;\chi =x,y 
\;\;, \\ [1ex] \label{I} 
{\cal E}(x,y)&:=&(x-y)v'(\frac{x+y}{2})
-v(x)+v(y)
\;\;, \end{eqnarray}  
in terms of coodinates $x,y$ and $v'(x):=\mbox{d}v(x)/\mbox{d}x$.  

This seems to be \emph{QT}! --  
The Eq.\,(\ref{Schroed}) looks like the  
\emph{quantum mechanical} von\,Neumann equation for a density operator $\hat f(t)$, 
considering $f(x,y;t)$ as its matrix elements.~\footnote{From here, one proceeds to find   
the probabilistic interpretation of \emph{QT} \cite{I09}.}
We recover the usual operator $\hat H$  
for the Hamiltonian function $H$. 
Yet an essential difference consists   
in the interaction $\hat {\cal E}$ between \emph{bra-} and \emph{ket-}states. 
Hilbert space and its dual (\emph{nota bene}) here are coupled by an unfamiliar 
\emph{superoperator}.~\footnote{How it disturbs the emergence of 
QT, has been discussed in various ways recently, cf. \cite{I09,tHooft06}.} 
 -- We find that: 
$$\;\hat {\cal E}\equiv 0\;\Leftrightarrow\;
\mbox{true potential}\;v\;\mbox{is constant, linear, or harmonic}\; .$$   
To appreciate this fact, we emphasize that we are here concerned with 
physics in continuum spacetime as a valid approximation.  
 
Different from a commonly used \emph{macroscopic potential} $V$, 
the \emph{true potential} $v$ becomes piecewise defined, 
when approaching smaller scales.~\footnote{Recall 
the random stacking of (1+1)-dimensional sheets mentioned in Section 2.3.} 
An arbitrarily differentiable $V$, especially, is an approximation to $v$ and  
differences between the two give rise to local 
fluctuations $\delta V$: 
%\begin{equation}\label{Ansatz} 
$v(\mathbf{x})=V(\mathbf{x})+\delta V(\mathbf{x})$.  
%\;\;. \end{equation} 
Following Section 2, there are two sources of \emph{fluctuations}: 
matter and spatiotemporal discreteness. 
  
Furthermore, there is an ``asymptotic freedom'' effect, due to   
spatiotemporal discreteness.~\footnote{The ``No Interaction Theorem'' can 
be illustrated as follows: 
\emph{Imagine each particle or 
wave-packet as a ``world tube''
in Minkowski space. In the center of mass frame, the two tubes
make an ``X''. The center of the X, where they meet, is the
interaction region. Now boost each particle to very high
energy. Because of Lorentz contraction, the tubes become so
flattened that the interaction region shrinks to less than a
Planck volume. Hence there is (very likely) no element of the
causet to represent this interaction region, whence there is no
interaction.} (R. Sorkin)} 

These remarks make it plausible to assume that the true potential  
$v$ is \emph{piecewise linear}, with pieces 
characterized by a typical ``linearity length'' $\delta$, with $\delta\gg l_{Pl}$, 
such that the continuum description is meaningful. 

Based on these considerations, it has been shown that the 
``Liouville equation'' (\ref{Schroed}) \emph{does} become the von\,Neumann equation of 
\emph{QT} \cite{I09}. -- Furthermore, the local fluctuations 
introduce a \emph{Lindblad term}, i.e., a natural 
decoherence and localization mechanism. This has been much looked for in recent 
studies, since it may lead to testable predictions. In particular, it would 
be a wellcome feat to derive the (non)existence of macroscopic 
``Schr\"odinger cat'' states from such an equation \cite{SchlosshauerZurekKiefer}.  

The inferred emergence of quantum mechanics here needs further study  
of the possibility and interpretation, or of the elimination, of 
negative probabilities, of entanglement, and with respect to the Born rule. 
Quantitative determination of  
induced fluctuations, e.g., $\delta V$, needs an     
understanding of the transition between microscopic spacetime-matter 
structure and emerging field theories, such as the Standard Model. 
Possibly, this requires a duality breaking mechanism. 

For this, it is  
essential to add to the two partial orders -- incorporating causality 
of spacetime, extensity of matter, and discreteness of both -- 
a model of their interactions, which invites further exploration.   
\begin{center} 
\begin{tabular}{p{5cm}}
\hline \phantom .
\end{tabular}
\end{center} 
$\mathcal{EPILOGUE:}$ \emph{``Nobody understands quantum mechanics.''} (R.P.\,Feynman) 
The arguments presented in this essay may open a new vista on this, ``The Problem''. 
A careful (re)examination of the fundamental notions of spacetime and matter may  
lead to the \emph{unforeseen possibility} to understand quantum mechanics by the 
methodology of and \emph{within} physics. \\ After all, \emph{``We can''}\,\,\ldots  
\vskip 2cm 
\noindent 
The author wishes to thank  
F.\,Berger, G.\,Kaufmann, L.\,Pesce and R.\,Sorkin for discussions or correspondence.  
%\vskip 12cm
%\eject
%%%%%%%%%%%%%%%%%%%%%%%%%%%

\end{document}